\documentclass[aps,twocolumn,superscriptaddress,groupedaddress,10pt,nofootinbib,floatfix]{revtex4}

\usepackage[utf8]{inputenc}
\usepackage{graphicx}  
\usepackage{dcolumn}   
\usepackage{bm}        
\usepackage{amssymb}   
\usepackage{amsmath}
\usepackage{epstopdf}
\usepackage[a4paper,margin=20mm]{geometry}
\usepackage{color}

\begin{document}

\title{Dynamic jamming of dense suspensions under tilted impact}

\author{Endao Han}
\thanks{These two authors contributed equally}
\affiliation{James Franck Institute and Department of Physics, University of Chicago, Chicago, Illinois 60637, USA}
\author{Liang Zhao}
\thanks{These two authors contributed equally}
\affiliation{James Franck Institute and Department of Physics, University of Chicago, Chicago, Illinois 60637, USA}
\author{Nigel Van Ha}
\affiliation{Swarthmore College, 500 College Ave, Swarthmore, PA 19081, USA} 
\author{S. Tonia Hsieh}
\affiliation{Department of Biology, Temple University, Philadelphia, PA 19122, USA}
\author{Daniel B. Szyld}
\affiliation{Department of Mathematics, Temple University, Philadelphia, PA 19122, USA}
\author{Heinrich M. Jaeger}
\email[E-mail: ]{h-jaeger@uchicago.edu} 
\affiliation{James Franck Institute and Department of Physics, University of Chicago, Chicago, Illinois 60637, USA}

\date{\today}

\begin{abstract}
Dense particulate suspensions can not only increase their viscosity and shear thicken under external forcing, but also jam into a solid-like state that is fully reversible when the force is removed. 
An impact on the surface of a dense suspension can trigger this jamming process by generating a shear front that propagates into the bulk of the system. 
Tracking and visualizing such a front is difficult because suspensions are optically opaque and the front can propagate as fast as several meters per second. 
Recently, high-speed ultrasound imaging has been used to overcome this problem and extract two-dimensional sections of the flow field associated with jamming front propagation. 
Here we extend this method to reconstruct the three-dimensional flow field.  
This enables us to investigate the evolution of jamming fronts for which axisymmetry cannot be assumed, such as impact at angles tilted away from the normal to the free surface of the suspension. 
We find that sufficiently far from solid boundaries the resulting flow field is approximately identical to that generated by normal impact, but rotated and aligned with the angle of impact. 
However, once the front approaches the solid boundary at the bottom of the container, it generates a squeeze flow that deforms the front profile and causes jamming to proceed in a non-axisymmetric manner.   
\end{abstract}

\maketitle


As non-Newtonian fluids, dense particulate suspensions show rich mechanical properties, including continuous shear thickening (CST) \cite{Wagner_PhysToday, Brady_1985, Cheng_2011} and discontinuous shear thickening (DST) \cite{Barnes_1989, Brown_Review_2014, Brown_JOR}, whereby their viscosities can increase dramatically under shear. 
In extreme cases, they can even transform into a jammed solid \cite{Nagel_Jamming, OHern, SJ} capable of supporting large applied stress, which ``melts'' and reverts back to the fluid state after the stress has been removed \cite{Scott, Ivo_Nature, Stone, Smith_NC, Bischoff}. 
Because of this property, dense suspensions have been applied as protection devices against impact or penetration, such as stabbing \cite{Wagner_Armor}. 
Another area of research where the stress response of dense suspensions is of interest is locomotion on, or through, terrain comprising fully liquid-saturated sand or mud, which is important for animals as well as wheeled or legged robots \cite{Goldman_Review,Mazouchova2013,Gravish,Li_Goldman_2013}. 

Currently, the transient stress response that determines the behavior during sudden application of mechanical loads and the associated transformation into a jammed solid are still much less understood than the dynamics of dense suspensions under steady-state forcing \cite{Wyart_Cates,Poon_Guy,Seto,Morris_2017}. 
Previous experiments have shown that jamming proceeds via fronts of localized, intense shear, which spread out from the point of forcing, rapidly propagate into the suspension, and change the suspension from a fluid to a solid-like state. 
Different types of forcing, such as impact, shear, and extension, were observed to generate similar dynamic jamming fronts \cite{Scott, EHan_NC, Brown_PRE, Ivo_Nature, Sayantan}. 
However, in all of these cases it was sufficient to perform two-dimensional (2D) imaging of the evolution of the associated flow field, given its axial or radial symmetry.  
When such symmetry is no longer assumed, such as during angled intrusions, the ability to track the flow field in all three dimensions becomes important.
This is challenging for dense suspensions, because typically they are optically opaque while a high frame rate is required to resolve the flow due to the fast front propagation speed. 
Last but not least, the front propagates deep into the bulk, so the scope of view needs to be large enough to capture the whole flow field.

Here we show how high-speed ultrasound imaging can be used to acquire three-dimensional (3D) flow fields in dense suspensions. 
For example, diffusion acoustic wave spectroscopy (DAWS) was developed to measure relative velocity and strain rate in suspensions \cite{Weitz_DAWS}. 
Combined with standard rheology, ultrasound speckle velocimetry has been used to measure steady flows in suspensions \cite{Manneville_2004, Manneville_2013}. 
Furthermore, recent experiments demonstrated that ultrasound is capable of imaging fast transient flows in dense suspensions \cite{EHan_NC, Sayantan}. 
These former experiments imaged 2D slices through the interior of a 3D suspension.
Here we extend this approach to extract the 3D flow field by stitching together spatially offset slices, each with a frame rate up to 10,000 s$^{-1}$.
We apply this method to image the transient flows inside dense suspensions that result from impact at different incident angles. 
This allows us to address the question to what extent these flows, and the associated jamming fronts, retain axisymmetry along the propagating direction and how they deform when approaching a solid boundary.

\section{Experimental methods}
\label{sec:EXPT}

\begin{figure*}[t!] 
\begin{center}
\includegraphics[scale=0.5]{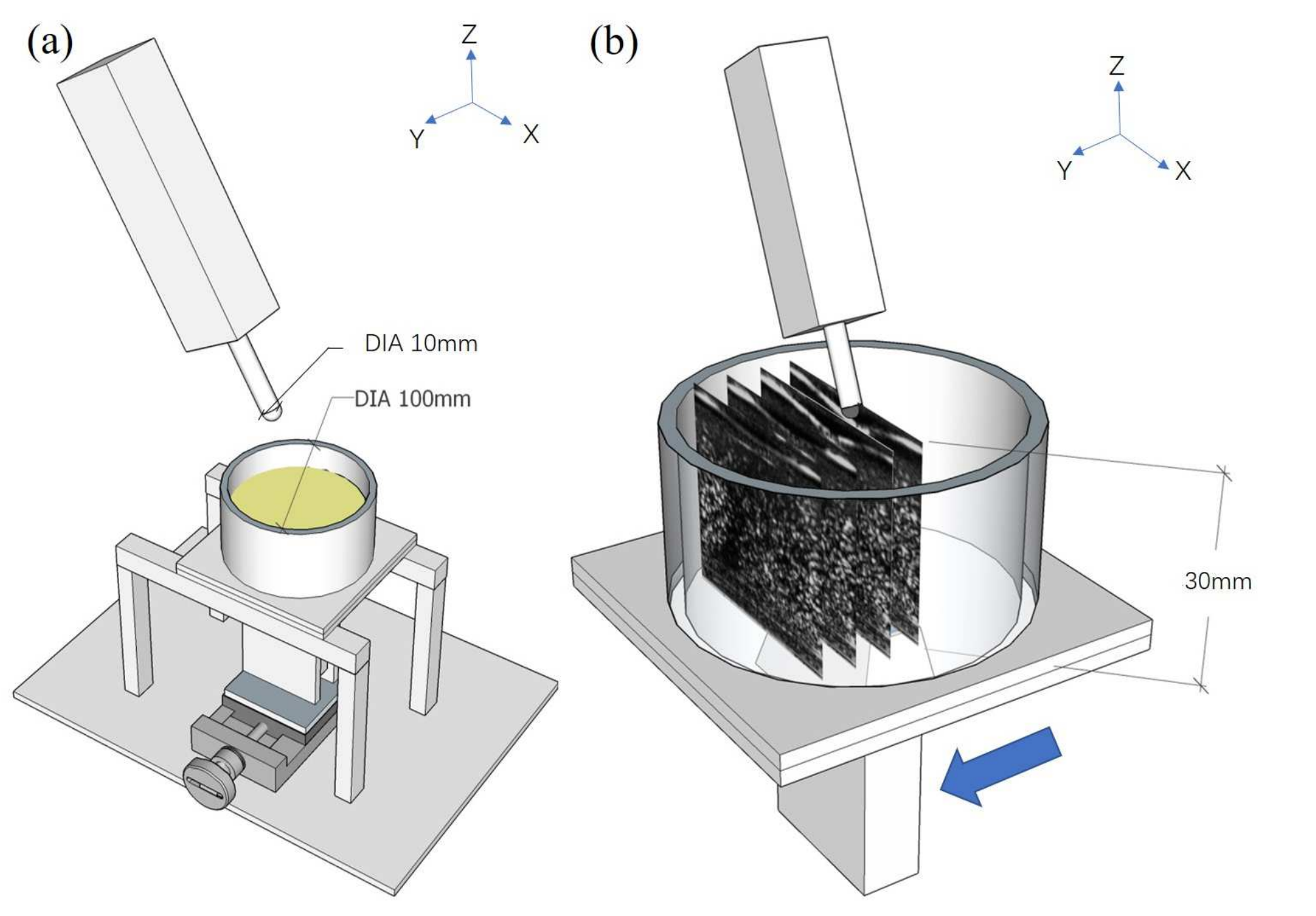}
\end{center}
\caption{Experimental setup. (a) The suspension sample is placed in a cylindrical container. An impactor motivated by an actuator is placed above the suspension. An ultrasound transducer is placed under the container, held by an optical stage. (b) Examples of the B-mode images obtained at different $y$ positions by moving the ultrasound transducer. Relevant dimensions of the setup are labeled in the figure.}
\label{Setup}
\end{figure*}

Our experiment studies impact at a suspension-air interface through ultrasound imaging and reconstruction of the resulting flow in three dimensions.
The experimental setup is schematically illustrated in Fig.~\ref{Setup}.
The suspension was contained in a cylindrical vessel with an inner diameter of 10~cm. 
The impactor was driven by a linear actuator (SCN5, Dyadic Systems) mounted above the container. 
The incident angle $\theta_\text{I}$\footnote{In cylindrical coordinates, this incident angle corresponds to a polar angle, which is normally represented by $\varphi$, and the notation $\theta$ normally represents the azimuthal angle. 
Since in literature $\varphi$ or $\phi$ usually represent the packing fraction of suspensions, in this paper we use $\theta$ to represent the polar angle.}, defined as the angle from the negative $z$ axis to the incident direction, and the impact speed $U_\text{p}$ were both adjustable. 
The plane in which the impactor rotated and moved was aligned with the center of the container, and we define this plane as the $y = 0$ plane. 
The head of the impactor was a hemisphere, so that it contacted the suspension surface in the same manner regardless of $\theta_\text{I}$. 
An ultrasound transducer (Philips L7-4) was placed underneath the bottom of the container and coupled to it acoustically through a layer of ultrasound gel. 
The transducer was a linear array of 128 piezoelectric elements aligned along the $x$ direction. 
Mounted on a translation stage, it could be moved to different $y$ positions with a resolution of 25~$\mu$m). 

The suspensions were prepared by dispersing dry cornstarch particles in an aqueous solvent. 
The solvent was made by mixing $398.0 \pm 0.1$~g cesium chloride (CsCl), $250.0 \pm 0.1$~g water and $250 \pm 0.1$~g glycerol. 
The density of the solvent was adjusted to match the density of the granules, which is 1.63~g/cm$^{3}$. 
The mass ratio  $\phi_\text{m} = m_\text{cs}/(m_\text{cs}+m_\text{sol})$ was 0.43, where $m_\text{cs}$ and $m_\text{sol}$ are the mass of cornstarch and the solvent, respectively. 
Air bubbles trapped in the suspension are strong scatterers of sound waves, which can significantly limit the penetration depth of the acoustic signal. 
Therefore, we de-bubbled the suspensions before imaging by placing them in a sealed container and shaking them at 3~Hz for two hours. 

In the experiments reported here, the impactor pushed into the surface of the suspensions with a constant speed set at $U_\text{p}=200$~mm/s. 
The impact angle $\theta_\text{I}$ was varied from 0$^\circ$ to 40$^\circ$. 
The ultrasound system was triggered and started to acquire 500 consecutive images at a frame rate of 4,000~s$^{-1}$ when the impactor reached a position 5~mm above the suspension surface (the Verasonics ultrasound system we used is capable of imaging up to 10,000 frames per second). 
Each acquisition generated a two-dimensional (2D) slice in the $x$-$z$ plane. 
For vertical impact, 2D images at $y=0$~mm (directly below the impactor) are sufficient to reconstruct the 3D flow field because of its rotational symmetry \cite{EHan_NC}. 
In order to visualize non-axisymmetric 3D flow under tilted impact, we moved the transducer along the $y$ axis as shown in Fig.~\ref{Setup} (b), and combined the 2D slices obtained from different $y$-positions. 
At each $\theta_\text{I}$, the transducer was moved from $y=0$~mm to $y=20$~mm, in increments of 5~mm. 
At each $y$, the impact experiment was repeated 3 to 9 times. 
After every impact, the suspension was fully relaxed by gently shaking and rotating the container. 
The accumulated data were used to reconstruct an averaged 3D flow field for each $\theta_\text{I}$.

\section{Impact-activated fronts}
\label{sec:Flow}

\begin{figure}
\begin{center}
\includegraphics[scale=0.5]{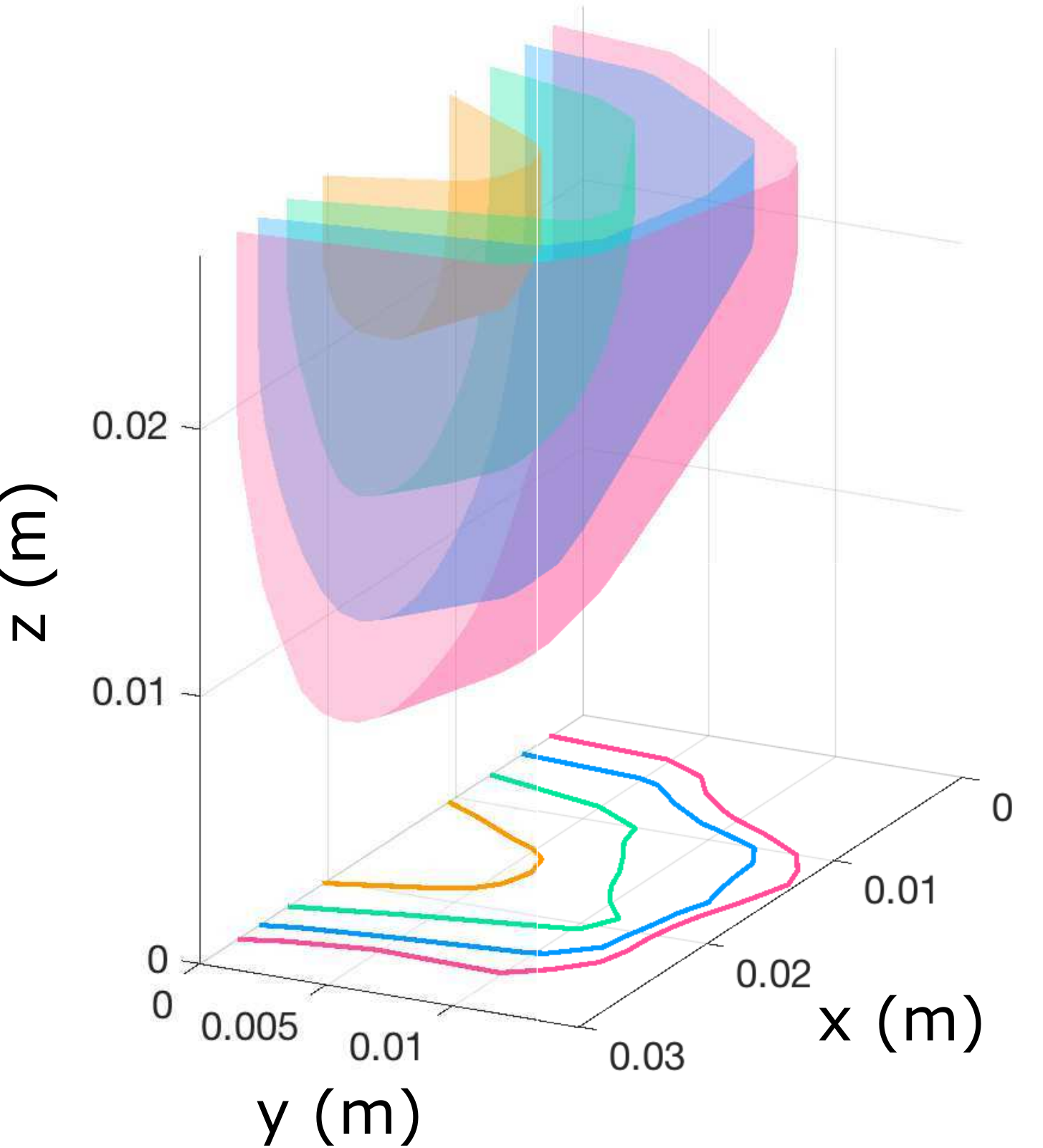}
\end{center}
\caption{Velocity isosurfaces under impact from $\theta_\text{I} = 10^\circ$. The orange, green, blue, and red surfaces are the isosurfaces of $v_\text{L} = 0.4 U_\text{p}$ at four consecutive time steps. The time difference between adjacent isosurfaces is 3.7~ms. The surface of suspension is at $z = 0.03$~m. Curves at the bottom outline the projections of these isosurfaces in the $x$-$y$ plane. }
\label{3D}
\end{figure}

To obtain the flow field, we used reconstructed B-mode images with trackable speckle patterns, whose motions represent the flow. 
A particle imaging velocimetry (PIV) algorithm was used to extract local velocities ${\bf v} = (v_x,v_z)$ from successive 2D images. 
We then calculated the longitudinal and transverse components $v_\text{L}$ and $v_\text{T}$ with respect to the incident direction as 
\begin{equation}
	\begin{split}
		& v_\text{L} = v_x \text{sin} \theta_\text{I} + v_z \text{cos} \theta_\text{I}, \\
		& v_\text{T} = v_x \text{cos} \theta_\text{I} - v_z \text{sin} \theta_\text{I}. 
	\end{split}
	\label{eq:vL_vT}
\end{equation} 
Fig.~\ref{3D} shows the velocity isosurfaces at $v_\text{L} = 0.4U_\text{p}$ at four different times, obtained from the reconstructed 3D flow. 
(See Supplemental Video 1 for the whole process.)
The impactor moved in the $y = 0$ plane, hitting the suspension surface at $z = 0.03$~m, and in this example, the incident angle was $\theta_\text{I} = 10^\circ$. 
Because the suspension thickened dramatically, the impactor did not penetrate significantly, but mainly caused a deformation of the suspension surface \cite{Scott}. 
Under the impactor, a transient flow developed in the form of a front, which propagated outward in all directions $x$, $y$, and $z$. 
It is this front that transforms the suspension from a fluid (ahead of the front) to a highly viscous, solid-like state (behind the front).

An impact event has two stages: The first stage is the free propagation regime, where the front is sufficiently far away from any solid boundary (bottom and side walls of the container). In this regime the front propagates much faster than the impactor speed \cite{EHan_NC}.  
In the example shown here it travels roughly 10 times as fast, i.e., it penetrates the whole depth of the suspension (3~cm) while the impactor pushes merely 3~mm down from the original surface position. 
As the front approaches the bottom, the already highly viscous, slowly deforming suspension behind the front is squeezed even more so that it turns into a jammed solid, ceases motion completely, and causes a dramatic increase in the force applied on the impactor \cite{Scott,Peters_2D}. 
We call this the interaction regime between the front and the bottom\footnote{In our experiments, the side wall of the container is further away (5~cm) from the impactor than the bottom (3cm), so we only consider the interaction with the bottom.}.

\begin{figure*}
\begin{center}
\includegraphics[scale=0.9]{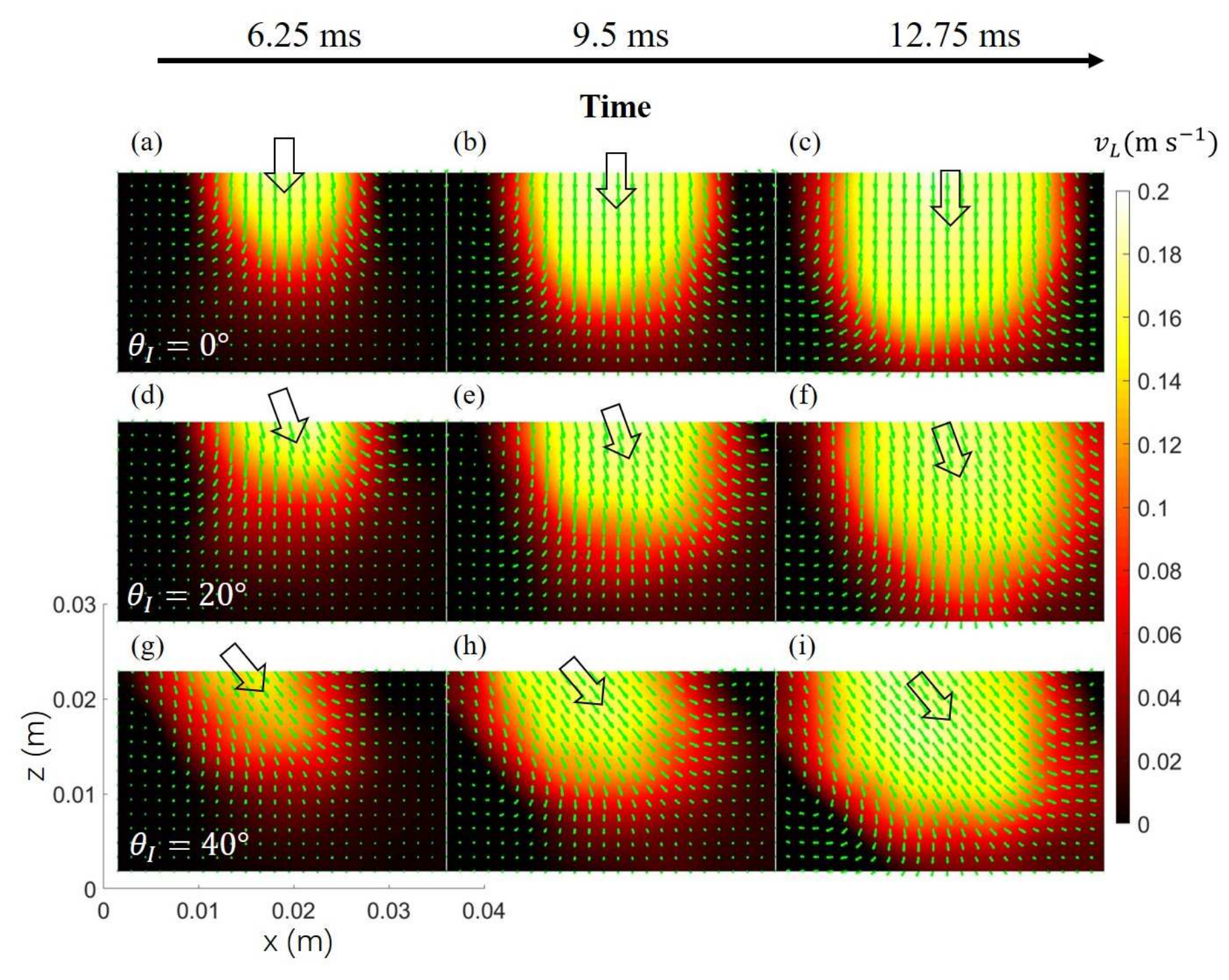}
\end{center}
\caption{Velocity field in the $x$-$z$ plane and isocontours for $v_\text{L}$ during impact at time $t$=6.25~ms (a,d,g), 9.5~ms (b,e,h), and 12.75~ms (c,f,i). The surface of the suspension and the bottom of the container are at $z=0.03$~m and 0~m, respectively. The green arrows label local velocities in the $x$-$z$ plane. The colors show $v_\text{L}$, with the upper limit (white) corresponding to $U_\text{p}$. The impactor hits the suspension at incident angles $\theta_\text{I} = 0^\circ$ (a-c), $\theta_\text{I} = 20^\circ$ (d-f), and $\theta_\text{I} = 40^\circ$ (g-i). The hollowed black arrows show directions of impact, with exaggerated penetration depths. }
\label{2DContour}
\end{figure*}

The reconstructed 3D flow field provides comprehensive visualization of this whole process, but most features are captured by the $x$-$z$ plane at $y=0$~mm. 
The flow fields in the $y = 0$~mm plane at three different angles $\theta_\text{I}=0^\circ$, $20^\circ$, and $40^\circ$ are presented in Fig.~\ref{2DContour}. 
Each row corresponds to a different $\theta_\text{I}$, and contains three succeeding time frames during the collision. 
In all these plots, the color map is based on the velocity component in the longitudinal direction $v_\text{L}$.

The top row in Fig.~\ref{2DContour} corresponds to impact normal to the suspension surface ($\theta_\text{I} = 0^\circ$). 
As mentioned above, the transient flow propagates in both longitudinal and transverse directions in the form of a front. 
In the bright region, the local velocities are almost all vertical, and their magnitudes are close to $U_\text{p}$, with very small gradients. 
In the dark region, the suspension flows at much slower velocities.
This leads to a sharp velocity gradient right on the edge of the bright region, which corresponds to both a large shear rate and a quick acceleration of the suspension. 
For Newtonian fluids, a larger shear rate leads to larger stress. 
However, shear thickening and shear jamming are controlled by stress instead of shear rate \cite{Ivo_Nature,Wyart_Cates}. 
Here, the suspension in the bright region has been turned into a highly viscous fluid by the impact-induced stress, as well as by reconfiguration of particles due to the accumulated strain \cite{EHan_NC}, and as a result, it is able to bear large stress at a small shear rate. 
According to experiments with a simpler geometry, in these bright regions the suspension not only shear thickens dramatically, but also evolves towards a shear jammed solid as the front keeps propagating \cite{EHan_PRF}. 
This is why we label such flows as impact-activated shear jamming fronts.

For oblique impact shown in Fig.~\ref{2DContour} (d)-(f) at $\theta_\text{I} = 20^\circ$ and (g)-(i) at $\theta_\text{I} = 40^\circ$, similar fronts are generated, though they are now tilted, lining up with $\theta_\text{I}$. 
In the rest of the paper, we will take a closer look at the similarities and differences between flows generated at different $\theta_\text{I}$. 
In Section~\ref{sec:FreeProp}, we will show that when the front is sufficiently far from the bottom, its shape is independent of $\theta_\text{I}$. 
In Section~\ref{sec:angleDependence}, we will map out the shear rate distribution in the flow, and use it to explain the anisotropy of the front.
In Section~\ref{sec:Boundary}, we will focus on the shear flow generated when the front approaches and finally interacts with the bottom, and discuss its effect on the jamming transition. 
All of these results reveal the importance of shear throughout the process of impact-activated solidification.

\section{Invariant front shape during free propagation}
\label{sec:FreeProp}


\begin{figure}
\begin{center}
\includegraphics[scale=0.6]{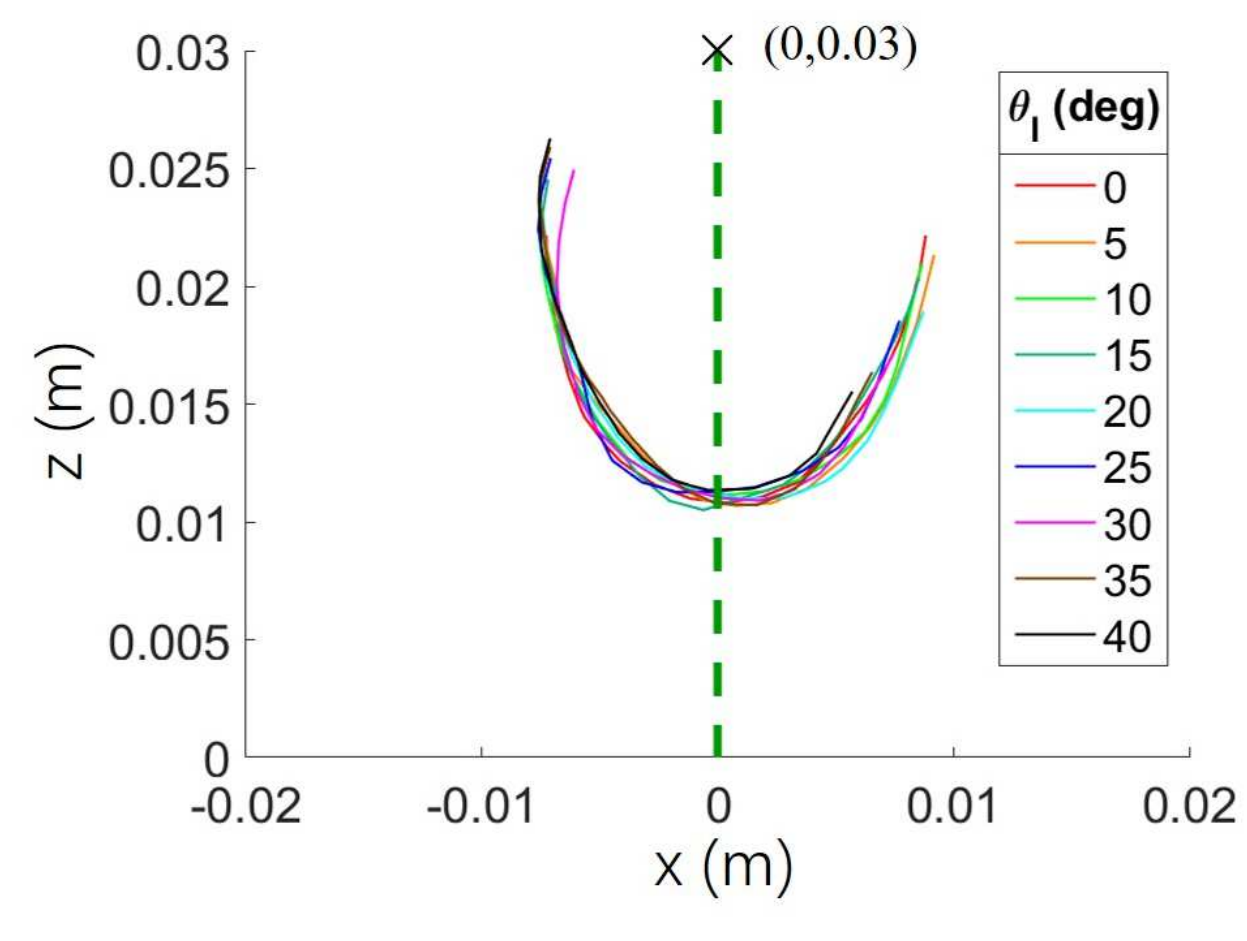}
\end{center}
\caption{Contours of $v_\text{L} = 0.4 U_\text{p}$ for $\theta_\text{I}$ ranging from $0^\circ$ to $40^\circ$, each rotated clockwise by $\theta_I$ accordingly. Each contour is obtained by averaging over three to nine experiments. The impactor hits the surface of the suspension at $x=0$~m, $z=0.03$~m, which is labeled by the cross, for all contours here after rotation. The dashed green line shows the impact direction. }
\label{Comparison}
\end{figure}

From Fig.~\ref{2DContour}, we can see that during free propagation the front is tilted along the incident direction of the impactor, while keeping a shape similar to that generated by upright impact. 
To demonstrate this more quantitatively, we rotate contours at $v_\text{L} = 0.4 U_\text{p}$ clockwise through the corresponding angle $\theta_\text{I}$, and plot results obtained from one exemplary frame in Fig.~\ref{Comparison}. 
(A comparison during the whole process is shown in Supplemental Video 2.)  
In this rotated frame, the impactor contacts the suspension surface at point $(x,z) = (0,0.03)$~m as labeled and pushes vertically down for every $\theta_\text{I}$. 
Within our experimental uncertainties, the contours are found to overlap well. 
As $\theta_\text{I}$ varies, the relative angle between the incident direction and gravity $\bf{g}$ changes, and the suspension surface curves in a different way. 
Nevertheless, they have very limited effect on the flow generated inside the suspension.
This means that to describe the free propagation of impact-activated fronts at different $\theta_\text{I}$, we can take full advantage of what we know about upright impact \cite{EHan_NC}.

\begin{figure} 
\begin{center}
\includegraphics[scale=0.6]{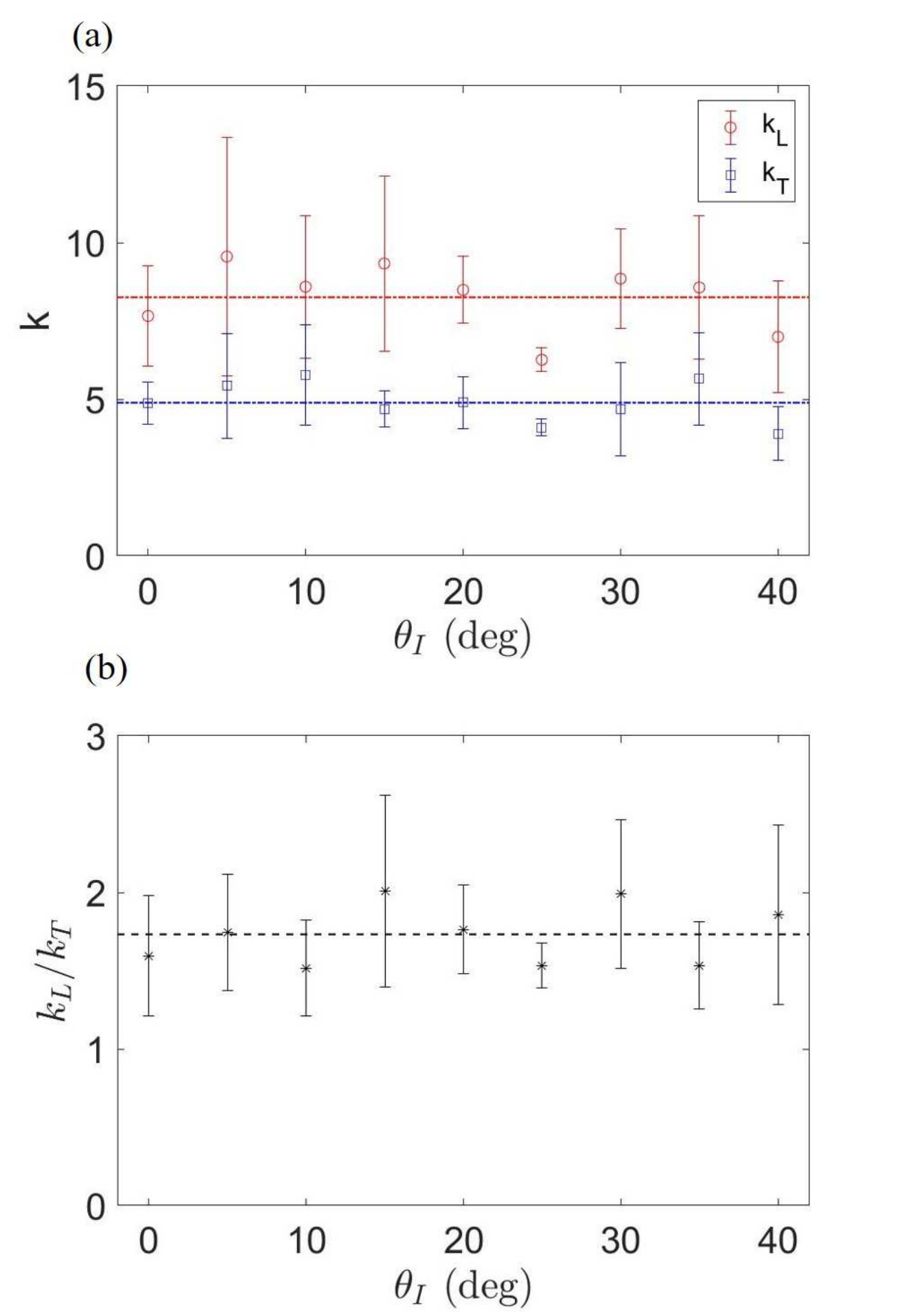}
\end{center}
\caption{(a) The $k_\text{L}$ and $k_\text{T}$ values (b) the ratio $k_\text{L} / k_\text{T}$ at different incident angles $\theta_\text{I}$. Error bars show the standard deviation from multiple measurements at each angle. The horizontal dot dash lines in (a) indicate the average values of $k_\text{L}$ and $k_\text{T}$, which are 8.3 and 4.9, respectively. The horizontal dashed line in (b) indicates $\sqrt{3}$. The overall standard deviations, using the data from all impact angles $\theta_\text{I}$, are 2.3 for $k_\text{L}$, 1.2 for $k_\text{T}$, and 0.42 for $k_\text{L}/k_\text{T}$. }
\label{Propagation Speed}
\end{figure}

To start with, we measured how fast the fronts propagated along the longitudinal and transverse directions.
Here we define the position of the front as the locations where $v_\text{L} = 0.4 U_\text{p}$. 
Based on the 2D flow fields in the $y = 0$~mm plane (Fig.~\ref{2DContour}), in each frame we plot out the isocontour that represents the front location (Fig.~\ref{Comparison}). 
With each contour, we then find its furthest reaching points in the longitudinal and transverse directions, and define their positions as the longitudinal and transverse front positions $R_\text{fL}$ and $R_\text{fT}$, respectively. 
Both $R_\text{fL}$ and $R_\text{fT}$ are linear functions of time and thus we obtain well defined front speeds $U_\text{fL}$ and $U_\text{fT}$. 
Lastly, we define dimensionless front propagation speeds $k_\text{L}$ and $k_\text{T}$ as 
\begin{equation}
	\begin{split}
		& k_\text{L} = U_\text{fL}/U_\text{p}-1, \\
		& k_\text{T} = U_\text{fT}/U_\text{p}, 
	\end{split}
	\label{eq:k}
\end{equation}
where the ``-1'' in $k_\text{L}$ is to subtract the impactor speed itself. 

Previous impact experiments at normal incidence \cite{EHan_NC, EHan_PRF} have shown that as $U_\text{p}$ increases, $k_\text{L}$ and $k_\text{T}$ each approach an asymptotic plateau that is independent of $U_\text{p}$. 
For the suspension we used here, $U_\text{p} = 0.2$~m/s was fast enough to be in this plateau regime.  
From the experimental measurements, neither $k_\text{L}$ nor $k_\text{T}$ show any apparent dependence on $\theta_I$, and their ratio $k_\text{L} / k_\text{T}\approx 1.7$ is a constant regardless of $\theta_\text{I}$, as shown in Fig.~\ref{Propagation Speed}. 
The origins of this anisotropic front propagation speeds are the finite accumulated strain required for shear jamming and the different types of shear in different directions, which will be discussed in detail in the following section.

\section{Shear rate distribution and anisotropy in front propagation}
\label{sec:angleDependence}

\begin{figure*}
\begin{center}
\includegraphics[scale=1.1]{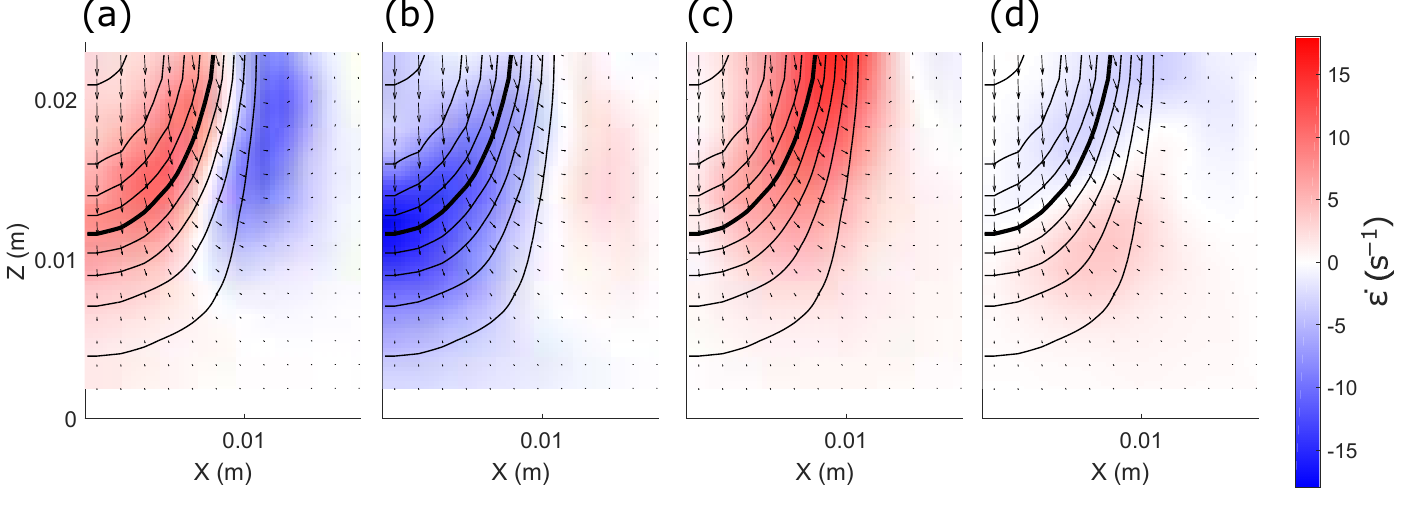}
\end{center}
\caption{Four components in the shear rate tensor (Eq.~\ref{eq:SR_tensor}). (a) $\partial v_r / \partial r$, (b) $\partial v_z / \partial z$, (c) $1/2 \cdot \partial v_z / \partial r$, (d) $1/2 \cdot \partial v_r / \partial z$. Black curves show contours from $v_z = 0.1 U_\text{p}$ to $v_z = 0.9 U_\text{p}$, with increment of $0.1 U_\text{p}$. The contour at $0.5 U_\text{p}$ is highlighted by the thick curves. }
\label{SRTensor}
\end{figure*}

With the flow field mapped out, we calculated the shear rate distribution in the system during free propagation. 
Given the invariance of the flow at different $\theta_\text{I}$, we can now focus on the case of the upright impact, where the system has a rotational symmetry in the azimuthal direction. 
In a system with rotational symmetry, the strain rate tensor $\dot{{\underline{\underline{\varepsilon}}}}$ can be written as 
\begin{equation}
    \dot{{\underline{\underline{\varepsilon}}}} = 
    \begin{bmatrix}
      \frac{\partial{v_r}}{\partial{r}} & 0 & \frac{1}{2}(\frac{\partial{v_r}}{\partial{z}}+\frac{\partial{v_z}}{\partial{r}}) \\
     0 & \frac{v_r}{r}  & 0 \\
      \frac{1}{2}(\frac{\partial{v_r}}{\partial{z}}+\frac{\partial{v_z}}{\partial{r}}) & 0 & \frac{\partial{v_z}}{\partial{z}} 
    \end{bmatrix}
. 
\label{eq:SR_tensor}
\end{equation}
Four of the components in this matrix are shown in Fig.~\ref{SRTensor} for an exemplary flow. 
Large shear rate focuses in a zone close to the front position, and the dominant term in Eq.~\ref{eq:SR_tensor} varies in different regions of the flow field. In the longitudinal direction, the dominant terms are the diagonal terms. 
The suspension is compressed in the vertical direction, and expands in the radial direction while keeping the total volume invariant, which is analogous to  pure shear in 2D. 
In the transverse direction, the dominant terms are the non-diagonal terms, especially $\partial v_z / \partial r$. 
In this case the local velocity changes along the $r$ direction, which is perpendicular to the direction of the velocities themselves (in the $z$ direction). 
This is analogous to simple shear in 2D, which is a combination of pure shear and a rotation.

\begin{figure*}
\begin{center}
\includegraphics[scale=1]{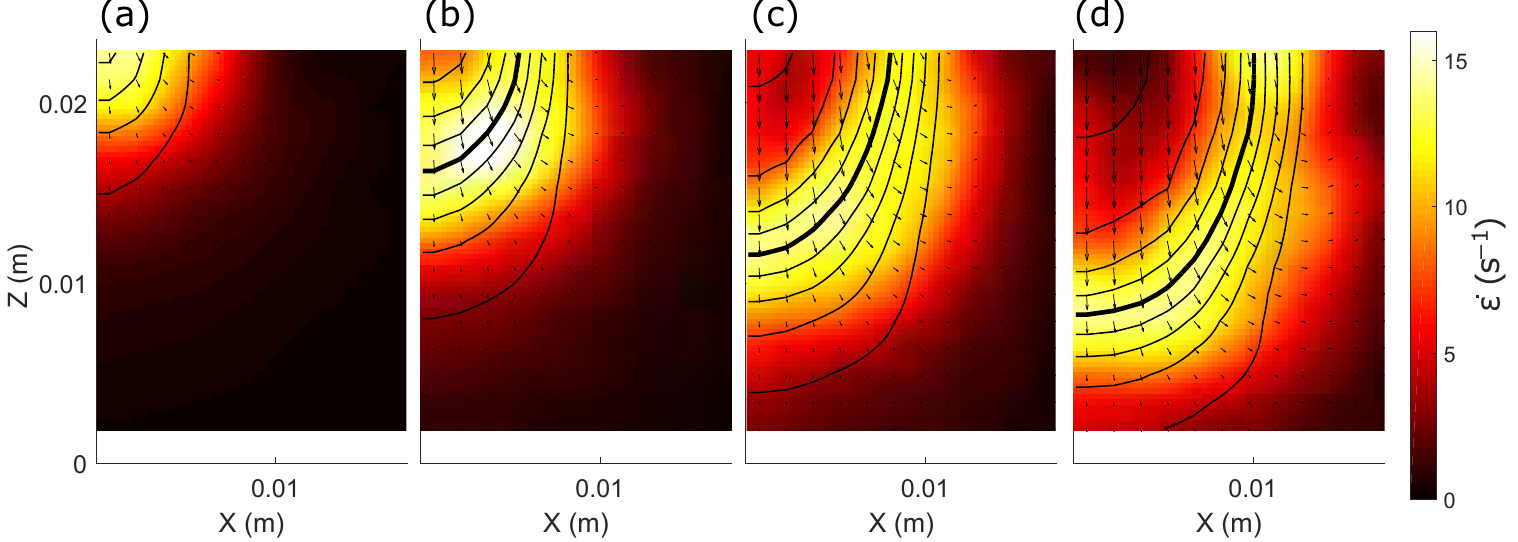}
\end{center}
\caption{Distribution of the scalar strain rate $\dot{E}$ at four different time points. From left to right are 2~ms, 4.5~ms. 7~ms, and 9.5~ms after the tip of the impactor reached the surface of the suspension. The contours are defined in the same way as Fig.~\ref{SRTensor}. Panel (c) and the flow shown in Fig.~\ref{SRTensor} are at the same time point. }
\label{SRIntensity}
\end{figure*}

\begin{figure}
\begin{center}
\includegraphics[scale=0.9]{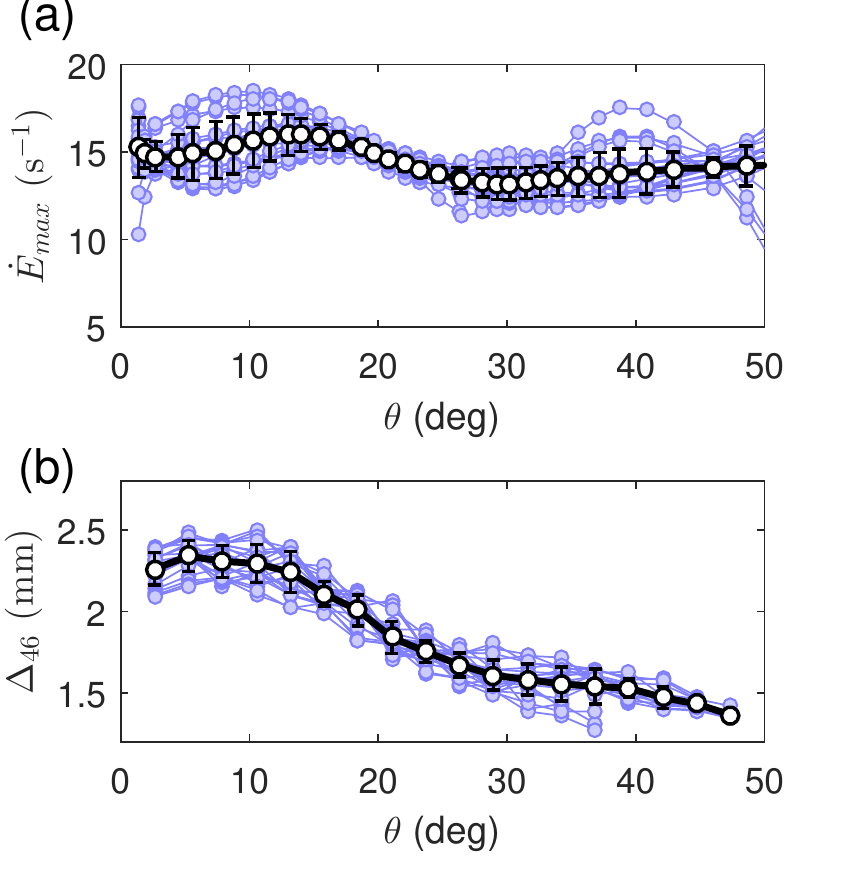}
\end{center}
\caption{Maximum strain rate $\dot{E}_\text{max}$ (a) at different polar angle $\theta$ with respect to the initial contact point of the impactor head and the suspension surface. Front width $\Delta_{46}$ (b), which is the distance between the contours of $v_z = 0.4 U_\text{p}$ and $v_z = 0.6 U_\text{p}$.}
\label{FrontWidth}
\end{figure}

To compare the shear rates in different regions, we introduce the strain rate scalar $\dot{E}$ defined by 
\begin{equation}
\dot{E} = \sqrt{ \left( \lambda_1^2+\lambda_2^2+\lambda_3^2 \right) / 2 }, 
\label{eq:Edot}
\end{equation}
where $\lambda_1$, $\lambda_2$, and $\lambda_3$ are the eigenvalues of $\dot{{\underline{\underline{\varepsilon}}}}$ (Eq.~\ref{eq:SR_tensor}) \cite{Seto_3D}. 
The spatial map of $\dot{E}$ at four different time points are shown in Fig.~\ref{SRIntensity}. 
As discussed above, $\dot{E}$ peaks very close to the front position where $v_z$ is approximately half of $U_\text{p}$. 
More importantly, $\dot{E}$ is almost invariant along the profile of the front, and it does not change with time as the front develops. 
In Fig.~\ref{FrontWidth}(a), we plot the maximum scalar shear rate $\dot{E}_\text{max}$ at different angles $\theta$, where $\theta$ is the polar angle defined in Fig.~\ref{AngleDemo} (See Appendix~\ref{app:Angle} for details).
The light blue data show the map of $\dot{E}_\text{max}$ from 5.5~ms to 10.5~ms after the impact started, and they show no significant variation over time. 
The open black circles are the average of all the blue data. We can see that $\dot{E}_\text{max}$ is a constant independent of the polar angle $\theta$.

The fact that $\dot{E}_\text{max}$ is angularly invariant leads to an interesting consequence, which is that the width of the front must change accordingly. 
As discussed above, the high shear rate zone is concentrated. 
Its width is a characteristic length scale of the flow, and we call it the front width $\Delta$.
For a simple estimation, the suspension accelerates from $v_z = 0$~m/s to $v_z \approx U_\text{p}$ as the front passes across. 
The characteristic shear rate $\dot{\varepsilon}$ in the high shear rate zone is $\dot{\varepsilon} \sim U_\text{p} / \Delta$ in the longitudinal direction $\theta = 0^\circ$, and $\dot{\varepsilon} \sim 1/2 \cdot U_\text{p} / \Delta$ in the transverse direction $\theta = 90^\circ$ (Eq.~\ref{eq:SR_tensor}). 
To compensate for the pre-factor $1/2$ in the non-diagonal terms and achieve an angle-independent shear-rate distribution, the front width $\Delta$ should be a function of the polar angle $\theta$: $\Delta(\theta = 0^\circ) \approx 2 \Delta(\theta = 90^\circ)$. 
To prove this, in Fig.~\ref{FrontWidth}(b) we plot out the distance between the contours at $v_z = 0.4 U_\text{p}$ and $v_z = 0.6 U_\text{p}$, which we call $\Delta_{46}$ (see details in Appendix~\ref{app:Angle}). 
Moving along the front profile, as $\theta$ changes from $0^\circ$ to $50^\circ$, $\Delta_{46}$ decreases from $2.4$ to $1.4$. 
Limited by the poor quality of the ultrasound signals obtained close to the suspension surface, we could not achieve measurements at higher $\theta$, but the trend is clear in Fig.~\ref{FrontWidth}.

Now let us revisit the anisotropy in the dimensionless front propagation speed $k$. 
Similar to the case of the front width, the ratio between $k_\text{L}$ and $k_\text{T}$ is roughly 2 as well. 
In previous work \cite{EHan_NC, EHan_PRF}, we have shown that in the high stress regime (fast impact), the speed with which a dense suspension shear jams is limited by having to build up the finite shear strain for rearranging the particles into a jammed configuration. 
We assume that this threshold strain is a scalar: when the suspension approaches this strain scale locally, its viscosity increases dramatically and develops towards a jammed solid. 
For a suspension that evolves towards jamming, the accumulated strain asymptotically adds up to 
\begin{equation}
    E_\text{J} = \left| \int_{-\infty}^{+\infty} \dot{E} dt \right|.  
	\label{eq:E}
\end{equation}
The scalar strain $E_\text{J}$,\footnote{For convenience, we only look at the absolute value of the integral.} the velocity vector ${\bf v}$, and the shear rate tensor $\underline{\underline{\dot{\varepsilon}}}$ are connected by Eq.~\ref{eq:SR_tensor}, Eq.~\ref{eq:Edot}, and Eq.~\ref{eq:E}.

As discussed above, the front propagation along the longitudinal (z axis) and transverse (radial direction, close to the surface) directions are the two most special cases. 
They each can be treated as a quasi-one-dimensional problem. 
Taking advantage of the invariant shape of the front during the propagation, we can write
\begin{equation}
	\begin{split}
		& v_z(z,t) = f_\text{L}(z + (k_\text{L}+1) U_\text{p} t), \\
		& v_z(r,t) = f_\text{T}(r - k_\text{T} U_\text{p} t),
	\end{split}
	\label{eq:velProfiles}
\end{equation}
where $f_\text{L}$ and $f_\text{T}$ are smooth functions. In the longitudinal direction, the front propagates towards $z \to -\infty$, and in the transverse direction, it propagates towards $r \to +\infty$. At a certain $z$ or $r$, they both approach $v_z = 0$ as $t \to -\infty$ and approach $v_z = -U_\text{p}$ as $t \to +\infty$. 
Consequently, we get
\begin{equation}
	\begin{split}
		\frac{D v_z(z,t)}{Dt} &= \frac{\partial v_z}{\partial t} + v_z \frac{\partial v_z}{\partial z} \\ 
		&= \left[ (k_\text{L}+1) U_\text{p} + v_z \right] \frac{\partial v_z}{\partial z}, \\
    	\frac{D v_z(r,t)}{Dt} &= \frac{\partial v_z}{\partial t} = -k_\text{T}U_\text{p} \frac{\partial v_z}{\partial r},
	\end{split}
	\label{eq:DvDt}
\end{equation}
for the longitudinal and transverse directions, respectively \cite{EHan_NC}. 
In the longitudinal direction, 
\begin{equation}
    \dot{E}_\text{L} = \frac{\sqrt{3}}{2} \cdot \frac{\partial v_z}{\partial z}, 
    \label{eq:Edot_L}
\end{equation}
and in the transverse direction, 
\begin{equation}
    \dot{E}_\text{T} = \frac{1}{2} \cdot \frac{\partial v_z}{\partial r} 
    \label{eq:Edot_T}
\end{equation}
(see Appendix~\ref{app:StrainRateIntensity}). 
Plugging Eq.~\ref{eq:Edot_L} and Eq.~\ref{eq:Edot_T} into Eq.~\ref{eq:E} and Eq.~\ref{eq:DvDt}, we get
\begin{equation}
	\begin{split}
		E_\text{JL} &= \left| \int_0^{-U_\text{p}} \frac{\sqrt{3}}{2} \cdot \frac{1}{(k_\text{L}+1)U_\text{p} + v_z} dv_z \right| \\ 
		&= \frac{\sqrt{3}}{2} \text{ln}\frac{k_\text{L}+1}{k_\text{L}} \approx \frac{\sqrt{3}}{2k_\text{L}}, \\
        E_\text{JT} &= \left| \int_0^{-U_\text{p}} \frac{1}{2} \cdot \frac{1}{k_\text{T}U_\text{p}} dv_z \right| = \frac{1}{2k_\text{T}}. 
    \end{split}
	\label{eq:E_LT}
\end{equation}
This means that the threshold accumulated strain to jamming is directly related to the dimensionless front propagation speed. 
Using our assumption that this threshold strain is the same, thus $E_\text{JL} = E_\text{JT} = E_\text{J}$, we have
\begin{equation}
    \frac{k_\text{L}}{k_\text{T}} \approx \sqrt{3} \approx 1.73, 
    \label{eq:kRatio}
\end{equation}
which agrees very well with our experimental measurements shown in Fig.~\ref{Propagation Speed}. 
We conclude that it is the difference in the type of shear along the moving shear front that leads to the anisotropy in the flow.

Predicting a precise value for this anisotropy is, however, more complicated. 
First, it depends on identifying the most appropriate parameter to measure the local shear strength in a 3D flow. 
In \cite{EHan_NC}, we proposed to use the scalar ``strain intensity'' $\mathcal{D}$ adapted from the Flinn diagram, which is often employed by geologists \cite{Ramsay}. 
Using $\mathcal{D}$ instead of $E$ for the data discussed here, the predicted anisotropy ratio becomes $k_\text{L} / k_\text{T} \approx 3/\sqrt{2} \approx 2.12$. 
This value is closer to the experimental results in Ref.\cite{EHan_NC}, but overestimates what we find in Fig.~\ref{Propagation Speed}. 
A second aspect to be taken into account is the shape of the impactor. 
Compared to flat-bottom impactors, for curved impactors the simple ratio of longitudinal and transverse components due to pure and simple shear, respectively, is likely to overestimate the anisotropy. 
Thus, while the shear front is still close to the impactor and for experiments with relatively shallow samples as discussed here, we would expect a smaller anisotropy in the front propagation than provided by the estimate based on either $\mathcal{D}$ or $E$. 
This may be the reason why we measure a 15\% smaller anisotropy with the hemispherical impactor.

\section{Interaction between front and solid boundary}
\label{sec:Boundary}

\begin{figure*}
\begin{center}
\includegraphics[scale=0.7]{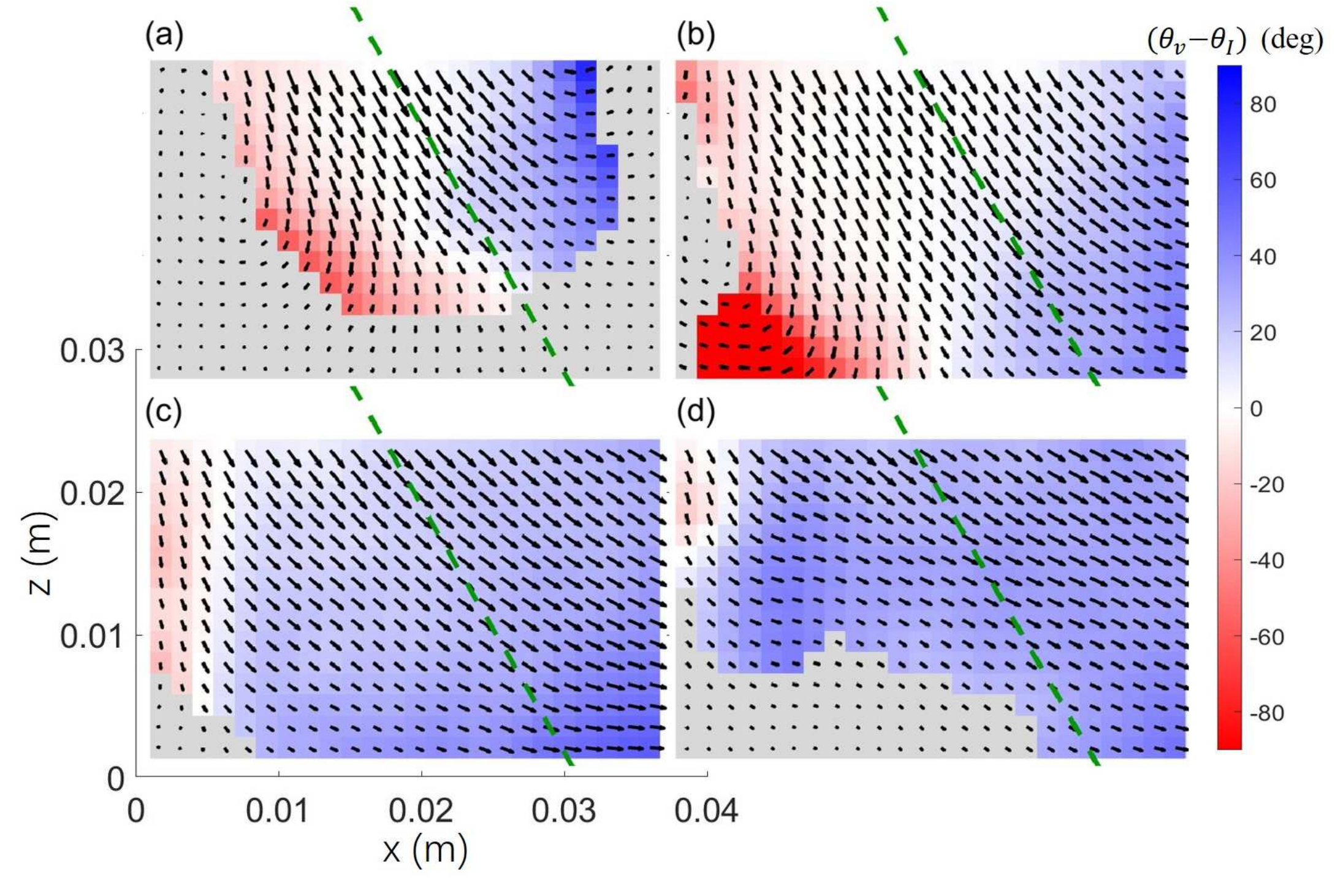}
\end{center}
\caption{Velocity directions for $30^\circ$ impact at $t$=9.5 (a), 16.25 (b), 22(c), 29 (d)~ms. Grids with a $v_L \geq 0.02$~m/s are colored red, blue or white according to $\theta_v-\theta_I$, as shown in the color map. Grids with $v_L \leq 0.02$~m/s are colored gray, which should be distinguished from the white region. The surface and bottom of the suspension are at $z=0.03$~m and 0~m. Dashed green lines are extension lines of the impact rod. The flow field here is obtained by averaging nine repeated experiments at the same $\theta_\text{I}$. }
\label{Boundary}
\end{figure*}

After free propagation, the impact-activated front starts to interact with the bottom of the container. 
Under upright impact, the jammed suspension forms a solid column that connects the impactor and the bottom, which can sustain a large amount of stress \cite{Scott,Peters_2D}. 
When it comes to tilted impact, does a similar column still extend between the impactor and the bottom, just tilted along the incident direction? 
This turns out not to be the case, due to more complicated flows generated while the front approaches the bottom. 
Furthermore, unlike during the free propagation stage, now the shape of the flow and how suspensions jam become dependent on $\theta_\text{I}$. 

To analyze this front-boundary interaction, we present the data in a different way. We pay attention to the velocity direction, and look at the angle $\theta_v$ between the velocity vector and negative z direction:  
\begin{equation}
	\theta_v = - \text{arctan}( v_x / v_z ). 
	\label{eq:theta_v}
\end{equation}
To be more precise, we look at the direction of the local velocities relative to the incident direction $\theta_v-\theta_\text{I}$. This is shown in Fig.~\ref{Boundary}, using $\theta_\text{I} = 30^\circ$ impact as an example. 
(See Supplemental Video 3 for the whole process.)  
Regions with positive $\theta_v-\theta_\text{I}$ are colored in blue, and negative values in red. 
The gray areas represent regions where the flow speed is below a threshold $|{\textit {\textbf v}}| < 0.1U_\text{p}$.

As Fig.~\ref{Boundary} (a) shows, when the front is far from the boundary, the red and blue regions are symmetric with respect to an axis through the impactor. 
This is consistent with the axisymmetric jamming front discussed above (e.g., in Fig.~\ref{Comparison}). 
Effects due to the solid boundary start to be observed when the distance between the front and the boundary becomes comparable to the front width $\Delta$, which is around 5~mm to 10~mm, as shown in Fig.~\ref{SRIntensity}. 
What happens subsequently is that the suspension near the bottom is squeezed by the front approaching from above and flows out to the sides. 
In upright impact experiments, we observed similar results, but symmetric horizontal flows are instead close to $z = 0$~m, with $\theta_v = \pm 90^\circ$. 
Tilted impact generates asymmetric squeeze flows, as shown in Fig.~\ref{Boundary}.
From left to right, the direction of the horizontal flow changes from $\theta_v = -90^\circ$ to $\theta_v = +90^\circ$, but this transition is not symmetric with respect to the dashed green line, i.e., the region where $\theta_v = \theta_\text{I}$ is no longer on the axis through the impactor. 
In Fig.~\ref{Boundary} (a) to (b) to (c), the boundary between red and blue keeps shifting to the left as the squeeze flow develops. 

At later times, as shown in Fig.~\ref{Boundary} (c), almost all velocities within the field of view have shifted to $\theta_v > \theta_\text{I}$. 
Comparatively, the suspension on the left side of the dashed green line flows slower, especially when it is close to the bottom.  
As the impactor pushes deeper, the bottom left part of the flow decelerates even more, and an almost static region with extremely small velocity ($v_\text{L} < 0.1U_\text{p}$) appears and develops. 
This is shown by the gray region in Fig.~\ref{Boundary} (d), while above it, the suspension is still moving at angles $\theta_v > \theta_\text{I}$. 
The static region continues to grow both vertically and horizontally until finally most of the suspension on the left of the dashed green line ceases to move. 
This is a general, robust phenomenon we find for all tilted impact, with larger tilt angles leading to larger $\theta_v$ in the end.

\begin{figure*}
    \begin{center}
        \includegraphics[scale=0.65]{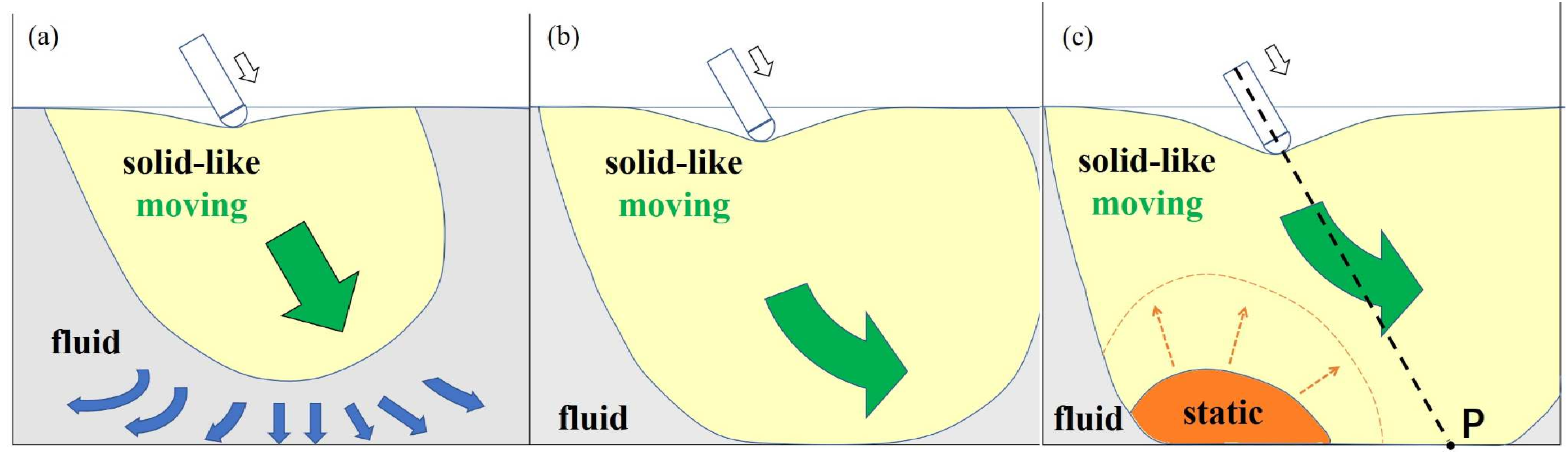}
    \end{center}
\caption{Schematic illustration of the flow when the front interacts with the solid boundary. (a)(b)(c) each corresponds to Fig.~\ref{Boundary} (b)(c)(d). The boundary between the solid-like (yellow) and fluid-like (gray) regions is the front. The green and blue arrows represent their motion, respectively. The static region is colored orange, and its expansion is shown by the dashed orange curve and arrows. The curved surfaces of the suspension are sketched based on the original ultrasound images.}
\label{Description}
\end{figure*}

Fig.~\ref{Description} schematically depicts how the solid boundary at the bottom modifies the flow and affects the jamming transition, based on the streamlines obtained from the flow field shown in Fig.~\ref{Boundary}. 
In this sketch, the highly viscous, solid-like region is colored yellow, and the fluid region gray. Their boundary is the contour $v_\text{L} = 0.4 U_\text{p}$ that represents the front position. 
In the solid-like region, the suspension moves along the green arrow that is parallel to the impact direction, and the local velocity is $(v_x, v_z) \approx ( U_\text{p} \text{sin} \theta_\text{I}, -U_\text{p} \text{cos} \theta_\text{I} )$. 
The blue arrows in Fig.~\ref{Description} (a) represent the squeeze flow between the front and the bottom. 
Because of this squeeze flow, the shear rates on the left and right of the front are no longer the same within 10~mm above the bottom. 
In the vertical direction ($z$), the velocity of the flow changes from $-U_\text{p} \cdot \text{cos} \theta_\text{I}$ to 0 while approaching the boundary, which applies for both sides. 
However, in the horizontal direction, the velocity difference on the left (dark red region in Fig.~\ref{Boundary} (b)) is larger than that on the right (blue region in Fig.~\ref{Boundary} (b)), because the direction of $v_x$ reverses on the left. 
Consequently, the left and right lobes of the front now accumulate strain at different rates. 
Since shear jamming of transient flow not only depends on the applied stress, but also the accumulated shear strain \cite{EHan_NC,EHan_PRF}, higher shear rate results in a shorter time to jam. 
We suspect that this is the reason why in Fig.~\ref{Boundary} the white boundary between red and blue keeps shifting to the left from the dashed green line, and also why the suspension ceases motion on the left, but keeps moving to the right as sketched in Fig.~\ref{Description} (b). 
In the final stage of the process shown in Fig.~\ref{Description} (c), the expansion of the static region (orange) forces the flow to circumvent this region, which leads to $\theta_v > \theta_\text{I}$.

Coming back to the question we posed at the beginning of this section: unlike for upright impact, when $\theta_\text{I} > 0^\circ$ the jammed (orange colored) region does not develop symmetrically with respect to the impact direction.
It is not initiated at point P in Fig.~\ref{Description} (c), which is where the dashed black line along the incident direction intersects with the bottom boundary. 
Instead, the formation of the solid-like region is initiated on the side of the line that is closer to the impactor. 
Its specific position on the $z = 0$ plane will depend on the size of the system, because when such a jamming front propagates, its dimension grows linearly with time, but its width $\Delta$ remains invariant \cite{EHan_PRF}. 
As discussed above, the effect of a solid boundary starts to appear when its distance to the front becomes comparable to $\Delta$. 
Thus, for a small system, the boundaries start to manipulate the flow immediately after the impact. 
For a larger system, however, the boundaries affect the flow only when the front has almost propagated across the whole system, and we expect the orange region in Fig.~\ref{Description} (c) to be initiated closer to point P.

\section{Conclusions}
\label{conclusions}

We demonstrated ultrasound as a facile high-speed 3D imaging technique to study dynamic shear jamming in dense suspensions under impact from different incident angles $\theta_\text{I}$. 
From the extracted flow fields, we find that the flow evolution can be separated into two stages: free propagation and boundary interaction. 
By visualizing jamming fronts generated at incident angles from $0^\circ$ to $40^\circ$, we obtained their propagation speeds and shapes. 
For a hemispherical impactor head, we find that the dimensionless front propagation speed in the longitudinal direction $k_\text{L}$ is approximately $1.7$ times the transverse propagation speed $k_\text{T}$, independent of the incident angle. 
This anisotropy can be explained quantitatively by considering the different types of shear in different regions of the flow. 
During the free propagation stage, fronts generated at different angles were shown to exhibit similar shapes. 
This implies that flow fields generated by impact at $\theta_\text{I}$ can be predicted by rotating the fields measured for upright impact. 
In the second stage, the front starts to interact with the solid boundary of the container. 
A squeeze flow is generated between the front and the boundary, which, for $\theta_\text{I}$ not equal to zero, produces an asymmetric shear rate on the sides of the curved front.
Consequently, the jammed solid that grows from the bottom to the surface of the suspension does not grow along the axis of impact. 
These observations agree with a scenario that explains impact-activated solidification as a dynamic shear jamming process, as previously applied to upright impact \cite{EHan_NC} as well as Couette shear \cite{Ivo_Nature} and quantitatively tested for simple shear in a quasi-1D configuration \cite{EHan_PRF}. 
However, still missing are more comprehensive 3D models that consider the tensorial forms of strain and stress and that can address more complex boundary conditions where the front interacts with obstacles and where the free surface is deformed by the impactor. 
Tilted impact into 3D suspensions as discussed here can provide an experimental system  to validate such models.

\section*{Acknowledgements}
We thank Ivo Peters for providing the PIV algorithm. We thank the referees for insightful suggestions and useful discussions. This work was supported by the US Army Research Office through grant W911NF-16-1-0078 and Chicago MRSEC (NSF) through grant DMR-1420709.  S.T.H. was supported by an NSF CAREER grant (IOS-1453106) and D.B.S. by NSF grants DMS-1418882 and DMS-1517100. We also acknowledge support from a ``Targeted Research Grant from Temple University'' (S.T.H. and D.B.S.) and from the Center for Hierarchical Materials Design (E.H.). 

\vspace{5mm}

\appendix

\renewcommand{\theequation}{A\arabic{equation}}
\setcounter{equation}{0}

\section{Representing strain rate with a scalar}
\label{app:StrainRateIntensity}

For simplicity, here we write Eq.~\ref{eq:SR_tensor} as
\begin{equation}
    \dot{{\underline{\underline{ \varepsilon }}}} =
	\begin{bmatrix}
        a & 0 & d \\
        0 & b & 0 \\
        d & 0 & c
    \end{bmatrix}, 
	\label{eq:simpleMatrix}
\end{equation}
where $a+b+c = 0$ because the suspension is incompressible. 
We write the eigenvalues of the matrix in Eq.~\ref{eq:simpleMatrix} as $\lambda_1$, $\lambda_2$, and $\lambda_3$, and sort them so that $|\lambda_1| \ge |\lambda_2| \ge |\lambda_3|$. 
Their values are 
\begin{equation}
    \begin{split}
        \lambda_1 &= b \\
        \lambda_{2,3} &= -\frac{b}{2} \pm \frac{1}{2} \sqrt{(a-c)^2+4d^2}. 
    \end{split}
\end{equation}
In this case, they each represent the strain rate along the corresponding principal axis. 
We can write the diagonal matrix $\text{diag} (\lambda_1, \lambda_2, \lambda_3)$ into the form 
\begin{equation}
    \dot{{\underline{\underline{ \varepsilon }}}} \sim
    \frac{2 \dot{E}}{\sqrt{3+4 \alpha^2}}
    \begin{bmatrix}
        1 & 0 & 0 \\
        0 & -(\frac{1}{2}+\alpha) & 0 \\
        0 & 0 & -(\frac{1}{2}-\alpha)
    \end{bmatrix},
	\label{eq:lambdaMatrix}
\end{equation}
where $\alpha = \frac{\sqrt{(a-c)^2+4d^2}}{2|b|} \in [0,1/2]$, and $\dot{E} = \sqrt{ (\lambda_1^2+\lambda_2^2+\lambda_3^2)/2 }$.  
For uni-axial compression in $z$ with isotropic flow in the $x$-$y$ plane, 
\begin{equation} 
    \dot{{\underline{\underline{ \varepsilon }}}} = 
    \begin{bmatrix}
        -\frac{1}{2} \frac{\partial v_z}{\partial z} & 0 & 0 \\
        0 & -\frac{1}{2} \frac{\partial v_z}{\partial z} & 0 \\
        0 & 0 & \frac{\partial v_z}{\partial z} 
    \end{bmatrix}. 
    \label{eq:Matrix_PureShear}
\end{equation}    
Thus $\alpha = 0$ and $\dot{E} = \sqrt{3}/2 \cdot \partial v_z / \partial z$.
For simple shear in the $x$-$z$ plane, we have 
\begin{equation}
    \dot{{\underline{\underline{\varepsilon}}}} =
    \begin{bmatrix}
        0 & 0 & \frac{1}{2} \frac{\partial v_z}{\partial r} \\
        0 & 0 & 0 \\
        \frac{1}{2} \frac{\partial v_z}{\partial r} & 0 & 0
    \end{bmatrix}. 
	\label{eq:Matrix_SimpleShear}
\end{equation}
In this case $\alpha = 1/2$ and $\dot{E} = 1/2 \cdot \partial v_z / \partial r$. 
Here we have ignored $\partial v_r / \partial z$ in the non-diagonal terms because it is much smaller than $\partial v_z / \partial r$ as shown in Fig.~\ref{eq:SR_tensor}.

\section{Front width and strain rate intensity}
\label{app:Angle}

\begin{figure}[b!]
    \begin{center}
        \includegraphics[scale=0.7]{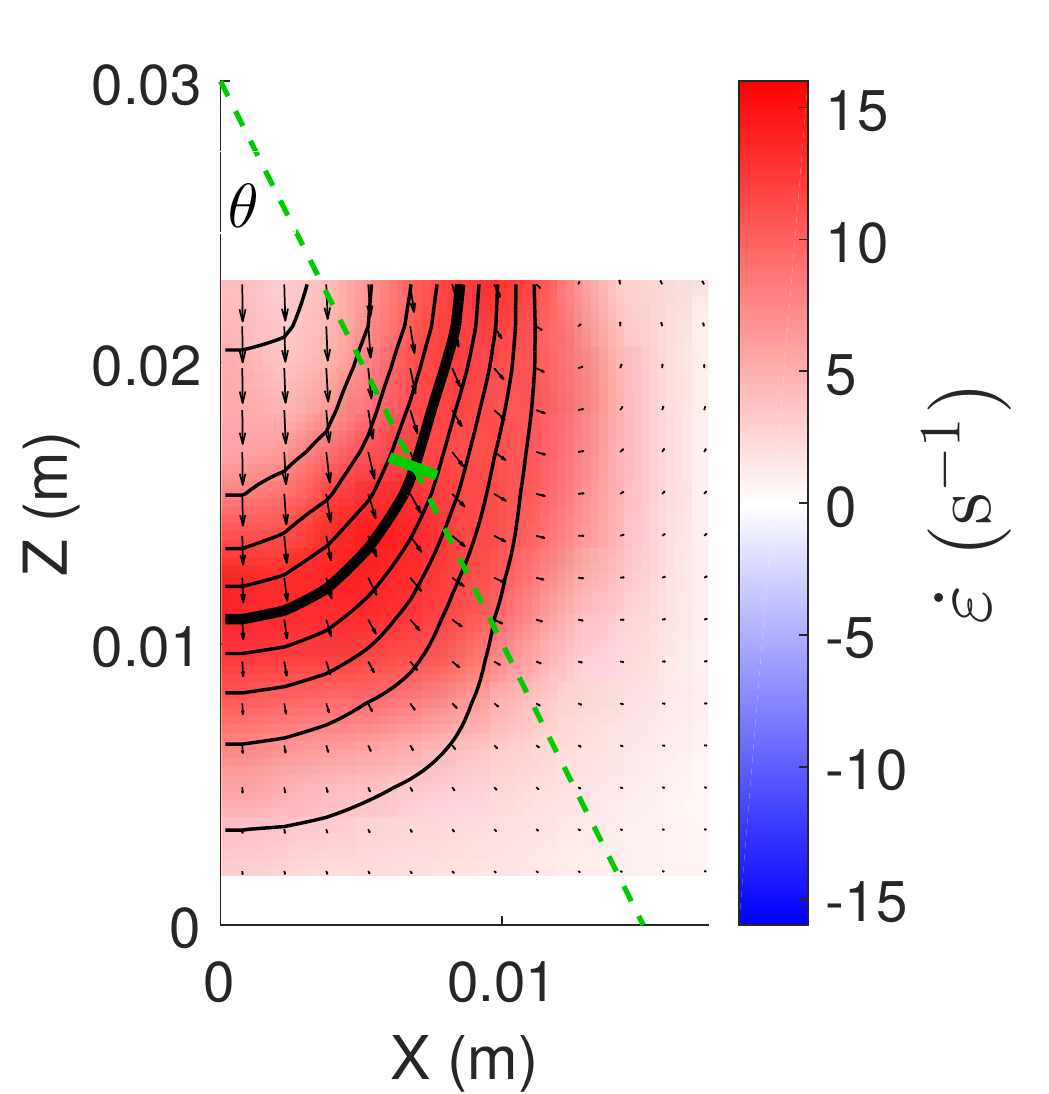}
    \end{center}
\caption{Demonstration of obtaining $\dot{E}_\text{max}$ and $\Delta_{46}$ at angle $\theta$. The color map shows $\dot{E}$. The dashed green line labels the direction at angle $\theta$. The length of the solid green line represents $\Delta_{46}$ at angle $\theta$. }
\label{AngleDemo}
\end{figure}

Here we demonstrate how we obtain $\dot{E}_\text{max}$ and $\Delta_{46}$ in Fig.~\ref{AngleDemo}. 
The green dashed line represents an arbitrary direction originates from the impact point at $x = 0$~m and $z = 0.03$~m. 
The corresponding polar angle $\theta$ is labeled. For each point in the flow, we can calculate its angle $\theta$, and obtain a distribution of local shear rate $\dot{E}$ as a function of $\theta$, ranging from small $\dot{E}$ close to zero (white areas) to values close to $\dot{E}_\text{max}$ (dark red areas). 
Then in each frame, we calculated the upper envelope of $\dot{E}(\theta)$, which gives us curves shown in Fig.~\ref{FrontWidth}(a). 
We define the front width $\Delta_{46}$ as the distance between the contour at $v_z = 0.4 U_\text{p}$ and $v_z = 0.6 U_\text{p}$ along the direction perpendicular to the contour $v_z = 0.5 U_\text{p}$. 
The corresponding line segment is shown by the solid green line in Fig.~\ref{AngleDemo}. It is perpendicular to the thick black curve, which represents $v_z = 0.5 U_\text{p}$ at their intersection point. 
$\Delta_\text{46}$ at angle $\theta$ is the length of this solid green line.

\section{Interactions between the ultrasound and the suspension}
\label{app:TransientDynamics}
\renewcommand{\theequation}{C\arabic{equation}}

Prompted by a comment by one of the reviewers, we briefly address whether the ultrasound itself might induce any jamming dynamics in the suspensions. 

Firstly, could the sound wave cause jamming by compressing the material locally? 
The ``dense'' suspension we used (0.43 mass ratio) was concentrated enough so that it could be jammed by shear, but it was still significantly below the isotropic jamming packing fraction (about 0.5 mass ratio). 
The bulk modulus of the liquid was about 4~GPa, and the bulk modulus of the saturated cornstarch particles was about 7 ~GPa \cite{Endao_SOS}. 
So under pressure, the liquid phase was compressed more than the solid phase, but not much. 

Now we estimate the upper limit of the acoustic pressure generated by the sound wave. 
The absolute maximum \textit{instantaneous} power that can be generated by our ultrasound system is 100~W per transducer element. 
While imaging, the power we used was much lower than this. 
The maximum \textit{average} power output per element of our machine was 8~W, and there were reflections at the interfaces, energy loss during propagation in the media, and a very low duty cycle that we used, all of which further decreased the power. 
Still, for a rough estimate we can calculate the pressure using 100~W per element. 
In a medium, the sound power $P$ along the propagation direction is 
\begin{equation}
P = \frac{Ap^2}{\rho c}, 
\label{eq:Power}
\end{equation}
where $A$ is the area of the surface, $\rho \approx 1600$~kgm$^{-3}$ is the density of our medium, $p$ is the pressure, and $c \approx 2000$~m/s is the speed of sound. 
The upper limit of the input power is $P = 100$~W/element $\times$ 128 elements $=12800$~W. 
Even if we assume that there is no energy loss and use the smallest area (the area of the transducer head, $35 \times 1$~mm$^2$), we would get $p \approx 3 \times 10^7$~Pa. 

When such a pressure is applied on GPa materials, the strain will be under $1\%$, and the relative change in volume will be less than $3\%$. 
This change will lead to an increase in packing fraction smaller than $0.01$. 
The actual change in our experiments should be orders of magnitude smaller than this. 
In comparison, to bring the system to jamming by compression, we would need to increase the packing fraction by about 0.07 or more. 
Therefore, the sound wave will not cause jamming by compression in our experiments. 

Secondly, does the ultrasound wave drives migration of the particles? 
We can think of two ways to generate relative motion between particles and the suspending liquid: one is due to scattering, the other is due to the inertia of the particles under high frequency oscillation (like shaking a cup of bubble tea). 
In our system the scattering is very weak. 
The wavelength of the ultrasound signal that we used was about $\lambda = 3$~mm. 
In comparison, the diameter of the cornstarch particles is approximately $d = 15$~$\mu$m, which is two orders of magnitude smaller than the wavelength. 
The cross section of a particle $\sigma$ is 
\begin{equation}
\sigma \propto \left( \frac{d}{\lambda} \right)^4 \cdot d^2, 
\label{eq:CrossSection}
\end{equation}
which means that $\sigma$ is more than $10^9$ smaller than the particle's actual area. 
As a result, the scattering is very weak, and the suspension behaves as a continuous medium when interacting with the ultrasound. 
The relative motion due to inertia is also highly limited. 
We carefully matched the density of the particles and the suspending solvent, so they should move the same way under the same acceleration. 
Combining these two factors, we think the relative motion between the particles and the liquid is negligible.

Lastly, how about jamming by shear? 
Our earlier work \cite{EHan_PRF} showed that to create a dynamic jamming front by shear, a local strain needs to be applied to shear the suspension out of the uniform, isotropic initial state, and bring the particles into contact networks. 
This requires a strain of order one that is applied along a particular direction. 
If instead of sustained unidirectional shear a brief oscillatory shear pulse is applied, as with the ultrasound, a stable contact network of particles cannot form, and thus no jammed state can be generated. 

Indeed, strong ultrasound can be used to manipulate grain structures \cite{Jia_unjam}, fluidize jammed granular bed \cite{Sebastien_unjam}, or apply a force that generates a shear wave inside soft elastic material (Acoustic Radiation Force Impulse (ARFI) elastography, for instance). For our system, we normally worried more about the signal being too weak instead of too strong. We never observed signs of ultrasound driven flow in dense suspensions with our setup. However, there are high power ultrasound systems, which could provide much stronger signals, and those might be powerful enough to manipulate the suspensions.


\end{document}